%% file: main.tex
\documentclass[10pt,twocolumn]{article}
\usepackage[margin=0.72in]{geometry}
\usepackage{amsmath,amssymb,booktabs,hyperref,url,tikz}
\usepackage{microtype}
\title{Tool Forge: A Validation-Carrying Toolchain for Governed Agentic Execution}
\author{Swanand Rao\\Next Moca Global, Inc.\\\texttt{swanand@nextmoca.com}\\\url{https://github.com/nextmoca/tool-forge}}
\date{May 2026}
\begin{document}
\maketitle
\begin{abstract}

Large language model agents are increasingly expected to perform operational work:
to call APIs, manipulate files, assemble workflows, and act inside enterprise
systems. Yet the tool layer on which this execution depends is still commonly
treated as either a hand-written integration artifact or a static list of
schemas exposed to a model. This paper introduces Tool Forge, a
validation-carrying toolchain for converting natural-language
capability intent into governed, sandbox-verified, cataloged tool artifacts and
for exposing those artifacts to agents through a token-efficient routing layer.

The central idea is that a tool should not be interpreted merely as executable
code. A production agent tool should be a capsule: a composition of intent,
capability contract, implementation, dependency policy, tests, documentation,
runtime validation evidence, lifecycle state, credential bindings, and routing
metadata. By making validation evidence part of the artifact, Tool Forge shifts
tool generation from model completion to software compilation. By routing over
compact catalog records rather than listing all schemas, it shifts agent tool
loading from static exposure to intent-scoped resolution.

We describe the conceptual model, architecture, validation pipeline, MCP-facing
Router, governance model, and initial reproducible benchmark results from the
current open-source implementation, available at \url{https://github.com/nextmoca/tool-forge}. Across 83 router
benchmark cases spanning lite, realistic, and adversarial suites, Tool Forge
Router achieves an aggregate micro-F1 of
0.908 while reducing estimated task-flow tool context by 99.49% relative to
naive full-catalog schema exposure. In a 25-case end-to-end generation probe
spanning L1 smoke, L2 realistic, and L3 adversarial local-tool tasks, the system
generates 25 of 25 tool bundles, reaches micro-F1 0.940 against deterministic
acceptance patterns, and passes 23 of 25 live sandbox validations. The
adversarial suites expose important remaining failure modes around negation,
semantically confusable tools, and edge-case validation, which we analyze as
open research and engineering problems. These results should be read as an
early systems benchmark for an open-source toolchain, not as a state-of-the-art
claim against other generators or agent frameworks. We argue that
validation-carrying tools and token-efficient routing are necessary systems
mechanisms for moving agentic systems from demos toward governed production
execution.

\end{abstract}
\section{Introduction}
\input{sections/01_introduction}

\section{Background and Motivation}
\input{sections/02_background}

\section{Tool Forge Conceptual Framework}
\input{sections/03_framework}

\section{System Architecture}
\input{sections/04_architecture}

\section{Contract and Evidence Model}
\input{sections/05_validation}

\section{Validation-Carrying Tool Capsules}
\input{sections/06_validation}

\section{Token-Efficient Tool Routing}
\input{sections/07_routing}

\section{MCP and Multi-Agent Orchestration}
\input{sections/08_mcp_orchestration}

\section{Experimental Protocol and Baselines}
\input{sections/09_methodology}

\section{Benchmark Evaluation}
\input{sections/09_benchmarks}

\input{benchmark_results}

\section{Governance, Isolation, and Enterprise Readiness}
\input{sections/10_governance}

\section{Limitations and Open Questions}
\input{sections/11_limitations}

\section{Conclusion}
\input{sections/12_conclusion}

\nocite{*}
\bibliographystyle{unsrt}
\bibliography{references}
\end{document}

%% file: sections/01_introduction.tex
Artificial intelligence systems are moving from a world of isolated inference
requests toward a world of persistent, tool-using, operational execution
entities. In the earlier phase of large language model adoption, the dominant
unit of interaction was the prompt. A user supplied context, a model returned
text, and the surrounding application interpreted the answer. The emerging
agentic pattern changes that assumption. The software layer above the model is
now expected to select tools, call APIs, transform files, update systems of
record, coordinate workflows, and preserve continuity across tasks [1, 2, 5,
7, 12].

This shift exposes a substrate problem. Agents can reason over a task, but they
can only act through tools. The quality, safety, and cost profile of an
agentic system therefore depends heavily on how tools are produced, validated,
cataloged, selected, and governed. A tool that is generated as a plausible
script is not necessarily an operational capability. It may have missing input
validation, stale dependencies, incomplete error handling, undocumented
credential requirements, unmocked network calls in tests, or no runtime evidence
that it performs the requested action. Conversely, a large tool catalog may be
semantically rich but operationally expensive if every schema is placed into the
model context for every task.

The common architecture still resembles explicit materialization. A tool is
constructed as code, then exposed as a full schema to the agent. When the
catalog grows, this approach couples capability growth to context growth. When
tools are generated, it couples model output to operational trust. Both
couplings are problematic. Context windows may grow, but token traffic remains
expensive, latency-sensitive, and cognitively noisy for the model. Code
generation may improve, but generated code remains software and must be treated
as untrusted until checked.

Tool Forge begins with a different interpretation. It treats a tool as a
validation-carrying capsule rather than as a bare function. The capsule contains
not only the executable implementation, but also the intent from which it was
derived, the capability contract it must satisfy, its dependencies, tests,
command-line interface, documentation, sandbox result, lifecycle state, and
router metadata. This makes a tool an inspectable operational artifact.

The second interpretation concerns tool exposure. Tool Forge Router treats the
agent-facing tool set as an intent-scoped projection over a governed catalog,
not as a complete schema dump. The agent first interacts with a small MCP
surface that can search, resolve, describe, and call tools. Full schemas are
loaded lazily only for the selected subset. This keeps the catalog large while
keeping the model-facing decision surface small.

The argument of this paper is twofold. The technical claim is that
validation-carrying tool artifacts and intent-scoped routing reduce two
practical bottlenecks in agentic systems: unverified tool generation and
tool-schema token bloat. The conceptual claim is that these mechanisms point to
a broader systems abstraction. In the same way that agent runtimes need control layers
for memory, identity, and orchestration [11, 13, 21], they also need a control layer
for governed executable capability.

This paper contributes: (1) a conceptual model of the validation-carrying tool
capsule; (2) a system architecture for compiling intent into governed Python
tool bundles; (3) a token-efficient MCP-compatible router for catalog-scale
tool selection; (4) a reproducible benchmark protocol with explicit routing
baselines and live-sandbox generation measurements from the current open-source
implementation; and (5) an analysis of limitations and open problems.

%% file: sections/02_background.tex
The modern tool-using agent is the result of several converging lines of work.
Large language models established broad natural-language reasoning and code
generation capabilities [1, 2, 3]. Tool-use methods such as ReAct [5],
Toolformer [6], Gorilla [7], ToolLLM [8], and HuggingGPT [9] showed that models
could plan over, select, and call external capabilities. Agent frameworks and
multi-agent systems such as AutoGen, MetaGPT, Reflexion, Voyager, and
Generative Agents explored persistent roles, collaborative execution, memory,
and feedback loops [11, 12, 13, 14, 15].

At the infrastructure layer, function calling and tool schemas became the
practical interface between model reasoning and software action. The Model
Context Protocol (MCP) generalizes this interface by defining a common protocol
for model applications to connect with tools, resources, and prompts [16]. MCP
is important because it gives agent systems a standard interoperability layer.
However, protocol-level interoperability does not solve the lifecycle problem of
where tools come from, how they are validated, how they are approved, or how a
large tool catalog is exposed without excessive context cost.

The issue is analogous to earlier transitions in software systems. Service
oriented architectures and microservices separated capabilities into callable
interfaces, but required service discovery, observability, deployment policy,
and operational ownership [24, 25, 26]. Container orchestration made units of
execution easier to deploy, but only because surrounding control planes tracked
state, health, scheduling, and policy [27, 28]. Modern data and ML systems
similarly learned that model artifacts require metadata, lineage, validation,
and reproducible pipelines [29, 30, 31].

Tool-using agents are now encountering the same pattern. The tool schema is not
the control plane. A generated integration must be reviewed, tested, sandboxed,
versioned, and governed. A third-party MCP server must be imported and
approved, not blindly trusted. A large catalog must be retrieved against intent,
not loaded wholesale. Without these surrounding mechanisms, tool use remains
closer to a demo capability than to an enterprise execution substrate.

Tool Forge is motivated by this operational gap. It is not primarily a better
prompt for generating Python code. It is a system that separates model-authored
synthesis from platform-enforced operationalization. This distinction is central
because the model may be cloud-hosted or local, strong or weak, deterministic or
variable. The trust boundary should not be the model response. The trust
boundary should be the validation and governance process wrapped around the
response.

%% file: sections/03_framework.tex
Tool Forge rests on a distinction between a tool as a materialized function and
a tool as an effective governed capability. In the first interpretation, a tool
exists once executable code has been produced. In the second interpretation, a
tool exists once the system has established a contract, an implementation, and
evidence that the implementation satisfies the contract within a declared
runtime boundary.

Let \(I\) denote a user intent. It may be an informal natural-language request
or a structured specification containing inputs, outputs, authentication
requirements, validation expectations, and live-sandbox values. Tool Forge maps
\(I\) into a capability contract:
\[
C = (n, P, K, O, R, H, E)
\]
where \(n\) is the tool name, \(P\) is the parameter set, \(K\) is the set of
required credentials and environment variables, \(O\) is the output contract,
\(R\) is the runtime class, \(H\) contains handling requirements for expected
failure modes, and \(E\) contains evidence extracted from user intent,
documentation, or indexed API material.

A Tool Forge tool capsule \(T\) is then:
\[
T = (C, X, D, U, V, G, A)
\]
where \(X\) is executable code, \(D\) is dependency policy, \(U\) is the user
surface such as CLI and README, \(V\) is validation evidence, \(G\) is
governance and lifecycle state, and \(A\) is audit and provenance metadata.
The capsule is not merely a packaging convention. It is the unit that the
catalog, router, and downstream agents reason about.

This model produces three useful invariants. First, a callable tool should have
a machine-readable schema and a dependency declaration. Second, a tool should
carry validation evidence rather than requiring the caller to infer trust from
the fact that code exists. Third, routing should operate over governed capsules,
not over raw functions.

The approach also changes the role of the model. The model is not the compiler.
It is a synthesis component inside a compiler-like pipeline. The platform owns
the intermediate representation, scaffolding, review gates, dependency policy,
sandbox execution, and catalog registration. This is especially important for
local small language models, which may be useful for privacy and experimentation
but should not be asked to author every operational artifact in a bundle.

Under this framework, the catalog is a set of tool capsule versions:
\[
\mathcal{K}=\{T_1,T_2,\ldots,T_N\}.
\]
A governance profile \(G_p\) defines which capsules may be resolved for a
particular agent, tenant, or workflow. A router receives an agent task \(q\) and
returns an intent-scoped session:
\[
S_q = R(q,\mathcal{K},G_p,k)
\]
where \(k\) is the maximum number of tools to expose. The session is a bounded
view over the catalog. Calls outside the session can be rejected. This makes the
agent's available tool set a runtime projection rather than a fixed global
list.

%% file: sections/04_architecture.tex
Tool Forge is organized around two loops: a generation loop and a routing loop.
The generation loop receives intent, synthesizes a capability contract, grounds
it in documentation where possible, generates code or a full bundle, applies
deterministic scaffolding, runs technical review, executes tests, and validates
the result in a sandbox. The routing loop indexes the catalog, applies
governance policy, resolves tools for a task, lazily exposes full schemas, and
audits calls.

The generation pipeline begins with intent enhancement only when needed. If the
user has already provided structured intent, enhancement is skipped because
unnecessary enhancement can introduce unrelated fields or providers. Capability
synthesis then produces a structured contract. For API integrations, Tool Forge
can extract documentation URLs from the prompt, index relevant pages, and derive
endpoint evidence such as HTTP methods, paths, provider vocabulary, and path
parameters. This grounding layer is conservative: it provides evidence for
reconciliation rather than absolute authority.

Generation recipes determine how much surface area the model owns. In
full-bundle mode, the model writes the tool implementation, CLI wrapper, tests,
README, tool card, requirements, and harness. This is appropriate for strong
cloud models when a richer initial bundle is desirable. In tiny mode, the model
writes only the core Python implementation. Tool Forge deterministically
creates the CLI, tests, metadata, README, requirements, and manifests. Local
model mode forces the effective recipe to tiny, reducing prompt size and
keeping weaker models away from brittle multi-file generation.

The validation stages are intentionally layered. Deterministic review catches
structural defects before sandbox execution. Unit tests must mock network calls
for external APIs. CLI help checks ensure that the command-line surface matches
the capability contract. Live sandbox validation then executes the tool with
runtime values. If required credentials or sandbox parameters are absent, the
backend pauses and asks for those values instead of pretending it can infer
them.

The routing loop is implemented by Tool Forge Router. The Router exposes a
small MCP-compatible surface: search tools, resolve tools, describe a selected
tool, call a tool, and list profiles. Generated Tool Forge tools and imported
third-party MCP tools can both be normalized into catalog cards. Imported MCP
tools begin in pending review and can be approved, blocked, pinned, or assigned
credential mappings. This lets Tool Forge act as a governed tool control plane
rather than a collection of one-off MCP servers.

The architecture supports a file-backed open-source mode and a hosted
system-of-record mode. In the file-backed mode, artifacts, router state,
profiles, governance records, sessions, and audit logs are inspectable on disk.
In a hosted deployment, Postgres and object storage can become the durable
system of record while preserving the same capsule and routing abstractions.

%% file: sections/05_validation.tex
The capability contract is the intermediate representation that separates
intent understanding from code generation. This intermediate layer is necessary
because natural-language intent is often underspecified, while executable code
is too concrete to be the first reliable representation. A contract can be
inspected, reconciled, repaired, and tested before the system commits to an
implementation.

The contract contains several classes of information. The first is the external
shape of the tool: name, description, input parameters, output fields, and CLI
flags. The second is the operational shape: required credentials, environment
variables, file outputs, runtime side effects, and validation inputs. The third
is the failure model: missing credentials, invalid parameters, provider errors,
network failures, rate limits, malformed responses, and permission failures.
The fourth is evidence: source documentation URLs, endpoint fragments, path
parameters, SDK method names, and user-provided examples.

This representation lets Tool Forge avoid two common errors. One error is
over-generation, where an enhancer or model invents inputs that were not part
of the requested tool. Another is under-specification, where a tool is generated
without an account identifier, parent folder identifier, authentication header,
or multi-step upload flow that the API actually requires. Contract
reconciliation does not eliminate those errors, but it creates a place where
they can be detected before code execution.

The evidence model is intentionally weaker than a formal specification. Most
public API documentation is prose, examples, and partial reference material,
not a machine-checkable theorem about provider behavior. Tool Forge therefore
uses documentation grounding as a source of high-impact evidence, not as a
source of unquestioned truth. If indexed documentation shows an endpoint such
as \texttt{/accounts/\{account\_id\}/folders/\{folder\_id\}/files/local\_upload},
the system can infer that both account and folder context may be important.
If the user prompt contains only a project identifier, the contract can be
flagged or amended before generation.

This evidence layer is also distinct from provider playbooks. A playbook is
useful when a flow is known and stable. Documentation grounding is useful when
Tool Forge encounters a new API and must extract enough structure to reduce
hallucination. The two approaches are complementary. The key architectural
requirement is that provider-specific knowledge must not become hidden
hardcoding in the core generator. The core should remain capable of discovering
evidence from arbitrary documentation, while playbooks can provide stronger
contracts for common integrations.

The contract-and-evidence model also makes regression analysis easier. When a
generated tool fails, the failure can be traced to a specific layer: intent
enhancement, capability synthesis, evidence extraction, contract reconciliation,
code generation, dependency policy, review, or sandbox execution. Without an
intermediate representation, those failures collapse into an opaque model
output.

%% file: sections/06_validation.tex
The defining property of a Tool Forge artifact is that validation evidence is
attached to the tool itself. A generated bundle contains the executable
implementation, a command-line wrapper, tests, a test harness, a README, a tool
card, runtime requirements, review findings, and sandbox results. These
artifacts create a record that can be consumed by humans, CI jobs, catalog
registries, and agents.

This design addresses a common failure mode in generated-code systems. A model
can produce code that looks plausible while missing edge cases, dependencies,
auth checks, or error handling. If the generated script is copied directly into
an agent runtime, the model response becomes the trust boundary. Tool Forge
replaces that boundary with a sequence of checks. The model may propose code,
but the platform decides whether the resulting capsule is admissible.

Dependency policy is one part of this process. Model-generated dependency pins
can be stale, over-specific, or incompatible with the execution environment.
Tool Forge normalizes dependency specifications through a policy layer so the
behavior is consistent across local and cloud modes. This prevents dependency
handling from becoming another prompt-level convention.

Documentation grounding is another part. Many production API integrations fail
not because the model cannot write Python, but because it uses the wrong
endpoint, misses an account or folder identifier, confuses API versions, or
omits a multi-step upload flow. Tool Forge's grounding layer attempts to extract
high-impact evidence from source documentation and reconcile it with the
capability contract. It is deliberately generic, not hardcoded to one provider.
Provider-specific playbooks can still exist, but the core system should not
require a hand-authored playbook for every new API on the Internet.

Validation evidence also improves downstream routing. A Router profile can
prefer sandbox-validated or approved tools. It can exclude failed, deprecated,
or blocked tools. It can expose only tools whose credential mappings are
available for a given tenant. In this sense, validation is not merely a
generation-time event. It becomes part of the runtime selection interface.

The capsule model does not prove semantic correctness. A tool can pass tests
and still be incomplete under production data, unusual permissions, rate
limits, or provider API drift. The claim is narrower and more practical:
validation-carrying capsules make the evidence explicit, machine-readable, and
available to the control plane.

%% file: sections/07_routing.tex
Tool routing addresses the consumption side of the lifecycle. A standard MCP
server can expose all tools through a tool-list response. This direct exposure
is simple and compatible, but it scales poorly for large catalogs. If there are
\(N\) tools and each full schema costs \(\bar{s}\) tokens, naive exposure costs
\(O(N\bar{s})\) tokens before the agent has begun the task.

Tool Forge Router changes the complexity profile. The agent first sees a small
stable MCP surface. The Router then filters the catalog by governance policy,
searches compact cards, creates an intent-scoped session, and reveals full
schemas only for selected tools. If compact cards cost \(\bar{c}\) tokens and
the session contains \(k\) tools, the conceptual cost becomes:
\[
O(N\bar{c}) + O(k\bar{s})
\]
with \(\bar{c} \ll \bar{s}\) and \(k \ll N\). If compact-card retrieval is
performed locally and only selected schemas are sent to the model, the
model-visible schema cost approaches \(O(k\bar{s})\).

This design is not merely an optimization. It changes the responsibility
boundary. A language model can choose from a list, but if the first operation is
to send a large list to the model, tool selection is already entangled with
token traffic, latency, and context noise. The Router performs retrieval and
policy filtering in ordinary software before the model sees the final decision
surface.

The Router can combine several implementation strategies behind this interface:
lexical scoring, semantic retrieval, curated profiles, lifecycle filters,
read/write classification, and future reranking. These are implementation
details beneath the control-plane abstraction. The paper's claim does not
depend on any single prefiltering technique; it depends on moving first-stage
selection and policy enforcement out of the model context and into a governed
software layer.

Anthropic's recent MCP optimization work identifies the same scaling pressure
from a complementary direction. Its advanced tool-use guidance argues that
agents should avoid stuffing all tool definitions into context and instead use
Tool Search for on-demand discovery, Programmatic Tool Calling for executing
multi-step tool workflows in code, and Tool Use Examples for improving
invocation accuracy beyond JSON schemas [38]. Its code-execution-with-MCP
pattern further observes that direct MCP calls make both tool definitions and
intermediate results flow through the model context; it proposes presenting MCP
servers as code APIs, using progressive disclosure or \texttt{search\_tools},
filtering and aggregating large results inside the execution environment, and
returning only the final useful output to the model [39].

Tool Forge adopts these methods as a model-agnostic control-plane strategy.
The Router's search, resolve, and describe operations correspond to deferred
tool loading and progressive disclosure. Tool cards, generated examples,
contract ledgers, and validation traces provide the usage evidence that schemas
alone cannot express. Router sessions provide the bounded set of tools that may
be used for a task. For data-heavy or multi-step workflows, the same catalog can
support a code-execution mode in which selected Tool Forge tools and imported
MCP tools are exposed as callable APIs inside a sandbox; the model sees the
orchestration code and final result rather than every intermediate tool
response. The distinction is that Tool Forge ties these token-saving techniques
to governance state, sandbox validation, credential mappings, and lifecycle
policy, rather than treating them only as inference-time compression.

The Router is MCP-compatible because its meta-tools are ordinary MCP tools. It
does not require a new wire protocol. It does require a retrieval-oriented
agent pattern: resolve tools for the task, inspect selected tools if needed,
and then call through the Router. For clients that require direct exposure,
selected profiles can mount tools directly, trading token efficiency for
compatibility.

%% file: sections/08_mcp_orchestration.tex
MCP compatibility gives Tool Forge an agent-facing protocol surface, but the
Router changes the operational role of an MCP server. Instead of requiring each
tool or small group of tools to be wrapped as a separate MCP server, Tool Forge
can serve as a governed aggregation layer over generated tools, human-authored
tools, and imported third-party MCP tools.

This is important for multi-agent systems. In a simple agent, a static tool list
may be acceptable. In a multi-agent environment, different agents need
different capability projections. A developer agent may require GitHub, CI, and
file-system tools. A revenue-operations agent may require CRM, spreadsheet, and
email tools. A document agent may require PDF, markdown, and storage tools. A
single global schema list forces all of these agents to share context and
policy surfaces even when their responsibilities differ.

Tool Forge profiles make capability surfaces explicit. A profile can define a
curated subset of tools, lifecycle requirements, credential mappings, and
imported MCP approvals. The Router then resolves task-specific sessions inside
that profile. This creates a two-level selection model: a human or platform
owner defines the permitted capability universe, and the Router resolves the
task-specific subset. The model is closer to capability-based security than to
unrestricted function calling [32, 33].

Imported MCP servers fit naturally into this model. A third-party MCP server
may expose dozens of tools, some read-only and some destructive. Tool Forge can
introspect or import the tool list, normalize each tool into a card, assign a
pending lifecycle state, and require explicit governance before exposure. This
lets organizations adopt MCP ecosystems without treating every imported server
as fully trusted by default.

The same mechanism also supports tool sessions. A session is a short-lived
binding between an agent intent and a resolved tool set. If an agent asks for
tools to upload a report and notify Slack, the session might include a PDF
converter, storage uploader, and Slack message sender. A later call to an
unrelated payment refund tool can be denied because it is outside the session.
This is a practical control against accidental tool drift during long reasoning
chains.

The orchestration implication is that agents do not need to carry the entire
tool universe in memory. They can ask the Router for the next relevant
capability surface. This is compatible with planning agents, workflow agents,
and human-in-the-loop systems. It also creates a clean integration point for
future HTTP MCP serving, MCP-server-from-intent generation, and enterprise
policy engines.

The current implementation is open source at \url{https://github.com/nextmoca/tool-forge}. The repository
includes the generator, Router, benchmark suites, API-document grounding
modules, dependency policy, local-model support, and self-hosting documentation.
This matters because the proposed control layer is intended to be inspectable
infrastructure rather than a closed demonstration.

%% file: sections/09_methodology.tex
The evaluation is designed around three research questions. RQ1 asks whether
intent-scoped routing can reduce model-visible tool context while preserving
tool-selection quality. RQ2 asks whether the generation pipeline can produce
complete tool bundles, not merely function bodies, that satisfy deterministic
artifact checks. RQ3 asks whether sandbox execution exposes failures that are
not captured by static artifact scoring.

The benchmark suite should be interpreted as an initial release benchmark. Its
purpose is to make the system's claims inspectable, reproducible, and
challengeable at open-source launch time. It is not a comprehensive production
certification suite and it is not presented as evidence of state-of-the-art
performance versus other frameworks. The results are most useful for evaluating
whether the proposed architecture has measurable signal and for identifying the
failure modes that should guide the next benchmark expansion.

For RQ1, the Router benchmark constructs synthetic but structured tool catalogs
and task intents. Each case defines a set of expected tools. The Router returns
selected tools for the intent, and the scorer computes true positives, false
positives, and false negatives. We report micro-precision, micro-recall,
micro-F1, macro-F1, average task-flow token estimates, and reduction versus a
naive full-catalog exposure baseline. Token estimates use the repository's
explicit approximation of characters divided by four. This approximation is
not tokenizer-identical, but it is deterministic and sufficient for comparing
relative exposure across fixed artifacts.

The Router benchmark's synthetic catalogs are intentionally controlled. This
enables repeatable measurement of catalog size, schema exposure, and confusable
tool families, but it does not fully capture the distribution of real
enterprise tool catalogs, naming conventions, permission structures, or
provider-specific edge cases.

The primary routing baseline is naive full-catalog schema exposure: every tool
schema in the catalog is made visible to the agent. This is a strong practical
baseline because it maximizes recall but scales linearly with catalog size and
pushes unrelated tools into the model context. We also record compact-catalog
and Router-meta-tool sizes. These are not competing agent policies by
themselves; they are exposure levels that clarify where token reduction enters
the system. The relevant comparison is between full schema exposure and the
Router's task-flow exposure after intent-scoped resolution.

For RQ2 and RQ3, the end-to-end generation benchmark uses fixed L1, L2, and L3
local-tool suites. Each case provides an intent, expected artifact patterns,
and sandbox inputs where applicable. The benchmark invokes Tool Forge, writes a
tool bundle, runs the generated harness, runs live sandbox validation, and then
scores the generated artifact against required and forbidden patterns. We
report generated-bundle count, precision, recall, F1, and sandbox pass count.
The E2E benchmark intentionally avoids live third-party APIs so that failures
measure generation and validation quality rather than provider availability.
This design makes the benchmark deterministic and inexpensive to reproduce, but
it also means the E2E result should not be generalized to arbitrary SaaS APIs,
multi-step OAuth flows, provider-specific upload protocols, or long-running
workflow automation.

We do not claim a matched superiority result against other tool-generation
systems in this version of the paper. Such a claim would require running the
same intents, constraints, sandbox inputs, and scoring harness against each
baseline. Instead, the paper reports two conservative comparisons that are
directly measured by the repository artifacts: Router exposure versus naive
full-catalog exposure, and deterministic artifact quality versus live sandbox
execution. This makes the current evaluation reproducible and falsifiable while
leaving cross-system comparisons to future benchmark work.

%% file: sections/09_benchmarks.tex
We evaluated the current open-source implementation using the initial offline
benchmark suites shipped in the repository. The reported numbers are intended
as transparent release measurements: they show the present behavior of Tool
Forge on fixed suites, and they define a baseline that future versions and
external contributors can improve or challenge. They are not a claim that Tool
Forge dominates other systems under a shared public benchmark.

The Lite suite contains 8 fast
regression cases over a 250-tool synthetic catalog. The L2 suite contains 50
realistic single-tool and multi-tool intents over a 600-tool catalog. The L3
suite contains 25 adversarial cases over a 500-tool catalog, including negation,
read/write ambiguity, and semantically confusable tool families.

The aggregate across all 83 Router cases is micro-F1 0.908 with 99.49% average
estimated task-flow token reduction relative to naive full-catalog exposure.
The strongest result is the L2 realistic suite, where Router selection reaches
micro-F1 0.958 and reduces task-flow tool context by 99.55%. The most useful
failure signal is the L3 suite, where micro-F1 drops to 0.786. Inspection of the
L3 failures shows weaknesses around negation and closely related tools: for
example, selecting file upload when the request asks for text-only Slack
notification, opening a pull request when the request asks for a GitHub issue,
or confusing upload, download, and delete operations in object storage.

These failures are important. They show that the Router's current retrieval
layer is already strong enough for large token reduction and many realistic
intents, but not yet sufficient as a sole safety mechanism for adversarial or
high-stakes tool selection. Governance profiles, mounted subsets, stricter
read/write classifiers, and second-stage semantic reranking are natural next
steps.

We also ran an end-to-end generated-tool benchmark over 25 fixed local-tool
cases: 10 L1 smoke tasks, 10 L2 realistic file and data-processing tasks, and 5
L3 adversarial safety tasks. Unlike the earlier catalog snapshot benchmark,
this run invokes the generator, writes full bundles, executes test harnesses,
runs live sandbox validation with benchmark-provided inputs, and then scores the
generated artifacts against deterministic acceptance patterns. The run generated
25 of 25 bundles with aggregate micro-F1 0.940, precision 1.000, recall 0.887, and
macro-F1 0.946. Live sandbox validation passed for 23 of 25 generated bundles.
Because these are local-tool cases rather than live third-party integrations,
the result measures bundle construction, local validation, dependency handling,
and sandbox behavior. It does not establish performance on providers such as
Slack, Stripe, Frame.io, Notion, GitHub, or Google APIs.

The per-tier generation scores are shown in the generated-tool results table.
L1 smoke tasks reached
micro-F1 0.985. L2 realistic tasks reached micro-F1 0.911 and were the hardest
portion of the run, primarily due to missing secondary return fields or
filesystem-error tests rather than wrong tool identity. L3 adversarial tasks
reached micro-F1 0.948 while preserving zero false-positive pattern matches.
Two live sandbox validations failed despite generated artifacts being written:
one CSV pivot case and one safe YAML case. We report these separately because
pattern coverage and live sandbox execution measure related but distinct
properties of generation quality.

%% file: benchmark_results.tex
\begin{table*}[t]
\centering
\small
\begin{tabular}{lrrrrrrr}
\toprule
Suite & Cases & Tools & Micro-F1 & Macro-F1 & Avg flow tok. & Naive tok. & Reduction \\
\midrule
Router Lite & 8 & 250 & 0.900 & 0.938 & 948.0 & 89665 & 98.94\\% \\
Router L2 realistic & 50 & 600 & 0.958 & 0.973 & 972.5 & 214464 & 99.55\\% \\
Router L3 adversarial & 25 & 500 & 0.786 & 0.780 & 811.1 & 178165 & 99.54\\% \\
\bottomrule
\end{tabular}
\caption{Tool Forge Router benchmark results. Token estimates use the repository benchmark convention of characters divided by four. Reduction compares the task-flow context against naive full-catalog schema exposure.}
\label{tab:router-results}
\end{table*}

\begin{table*}[t]
\centering
\small
\begin{tabular}{lrrrrrr}
\toprule
Suite & Tools & Full schema tok. & Compact tok. & Router tok. & Avg flow tok. & Flow reduction \\
\midrule
Router Lite & 250 & 89665 & 30130 & 284 & 948.0 & 98.94\\% \\
Router L2 realistic & 600 & 214464 & 72222 & 284 & 972.5 & 99.55\\% \\
Router L3 adversarial & 500 & 178165 & 60058 & 284 & 811.1 & 99.54\\% \\
\bottomrule
\end{tabular}
\caption{Context exposure baselines for routing. Full schema tokens estimate naive exposure of every tool schema; compact tokens estimate catalog-card exposure; Router tokens estimate the stable Router MCP surface; average flow tokens estimate the task-scoped selected-tool context.}
\label{tab:context-exposure-baselines}
\end{table*}

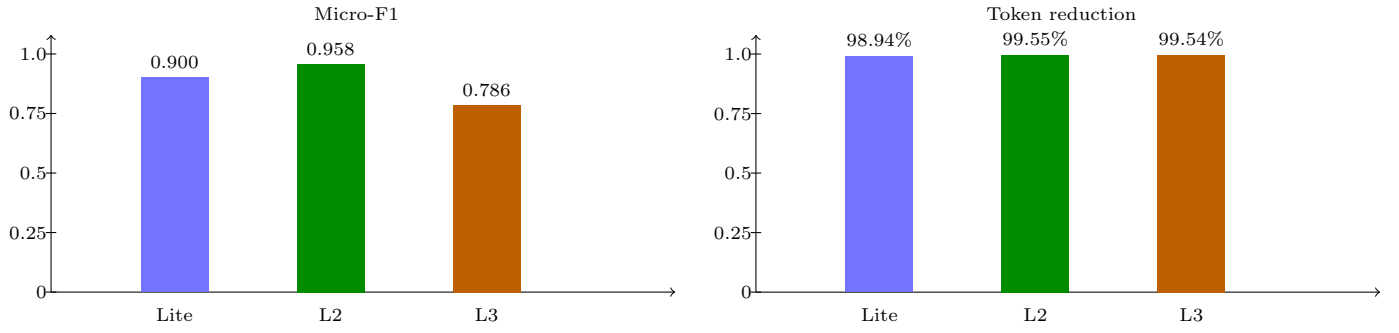
\begin{figure*}[t]
\centering
\begin{minipage}{0.48\textwidth}
\centering
\begin{tikzpicture}[x=1.72cm,y=3.15cm]
\draw[->] (0,0) -- (4.8,0);
\draw[->] (0,0) -- (0,1.08);
\foreach \y in {0,0.25,0.5,0.75,1.0} {\draw (-0.04,\y) -- (0.04,\y) node[left,font=\scriptsize] {\y};}
\fill[blue!55] (0.69,0) rectangle (1.21,0.9000);
\node[below,font=\scriptsize] at (0.95,-0.03) {Lite};
\node[above,font=\scriptsize] at (0.95,0.9000) {0.900};
\fill[green!55!black] (1.89,0) rectangle (2.41,0.9577);
\node[below,font=\scriptsize] at (2.15,-0.03) {L2};
\node[above,font=\scriptsize] at (2.15,0.9577) {0.958};
\fill[orange!75!black] (3.09,0) rectangle (3.61,0.7857);
\node[below,font=\scriptsize] at (3.35,-0.03) {L3};
\node[above,font=\scriptsize] at (3.35,0.7857) {0.786};
\node[font=\scriptsize] at (2.35,1.17) {Micro-F1};
\end{tikzpicture}
\end{minipage}
\hfill
\begin{minipage}{0.48\textwidth}
\centering
\begin{tikzpicture}[x=1.72cm,y=3.15cm]
\draw[->] (0,0) -- (4.8,0);
\draw[->] (0,0) -- (0,1.08);
\foreach \y in {0,0.25,0.5,0.75,1.0} {\draw (-0.04,\y) -- (0.04,\y) node[left,font=\scriptsize] {\y};}
\fill[blue!55] (0.69,0) rectangle (1.21,0.9894);
\node[below,font=\scriptsize] at (0.95,-0.03) {Lite};
\node[above,font=\scriptsize] at (0.95,0.9894) {98.94\%};
\fill[green!55!black] (1.89,0) rectangle (2.41,0.9955);
\node[below,font=\scriptsize] at (2.15,-0.03) {L2};
\node[above,font=\scriptsize] at (2.15,0.9955) {99.55\%};
\fill[orange!75!black] (3.09,0) rectangle (3.61,0.9954);
\node[below,font=\scriptsize] at (3.35,-0.03) {L3};
\node[above,font=\scriptsize] at (3.35,0.9954) {99.54\%};
\node[font=\scriptsize] at (2.35,1.17) {Token reduction};
\end{tikzpicture}
\end{minipage}
\caption{Router benchmark trends. L2 realistic tasks preserve high selection quality while L3 adversarial tasks expose semantic-confusion failure modes. All suites preserve large estimated reductions in task-flow tool context versus naive full-catalog exposure.}
\label{fig:router-benchmark-graphs}
\end{figure*}

\begin{table}[t]
\centering
\small
\begin{tabular}{lrrrrrr}
\toprule
Suite & Cases & Gen. & Sandbox & Micro-F1 & Prec. & Recall \\
\midrule
E2E L1 smoke & 10 & 10 & 10/10 & 0.985 & 1.000 & 0.971 \\
E2E L2 realistic & 10 & 10 & 9/10 & 0.911 & 1.000 & 0.836 \\
E2E L3 adversarial & 5 & 5 & 4/5 & 0.948 & 1.000 & 0.902 \\
\bottomrule
\end{tabular}
\caption{End-to-end generated-tool benchmark over fixed L1/L2/L3 local-tool suites. ``Sandbox'' reports live sandbox validations passed out of generated bundles; pattern scores are deterministic artifact checks.}
\label{tab:generated-results}
\end{table}

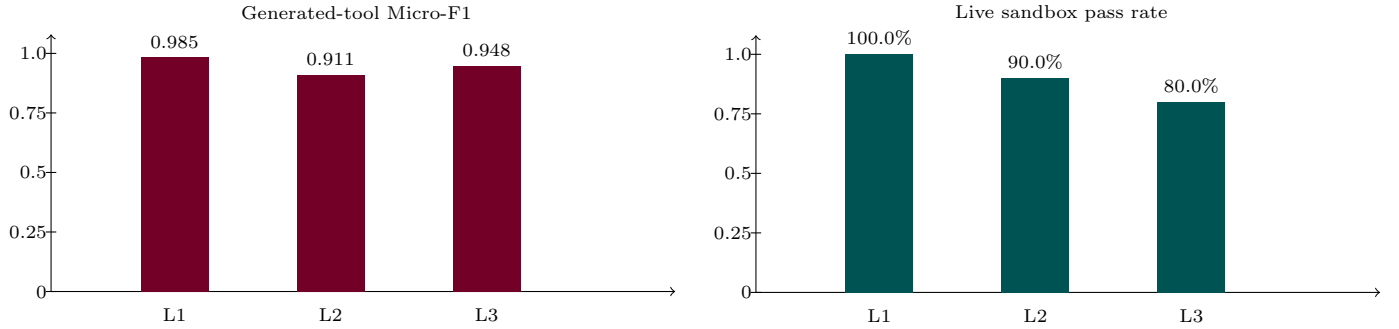
\begin{figure*}[t]
\centering
\begin{minipage}{0.48\textwidth}
\centering
\begin{tikzpicture}[x=1.72cm,y=3.15cm]
\draw[->] (0,0) -- (4.8,0);
\draw[->] (0,0) -- (0,1.08);
\foreach \y in {0,0.25,0.5,0.75,1.0} {\draw (-0.04,\y) -- (0.04,\y) node[left,font=\scriptsize] {\y};}
\fill[purple!60!black] (0.69,0) rectangle (1.21,0.9851);
\node[below,font=\scriptsize] at (0.95,-0.03) {L1};
\node[above,font=\scriptsize] at (0.95,0.9851) {0.985};
\fill[purple!60!black] (1.89,0) rectangle (2.41,0.9106);
\node[below,font=\scriptsize] at (2.15,-0.03) {L2};
\node[above,font=\scriptsize] at (2.15,0.9106) {0.911};
\fill[purple!60!black] (3.09,0) rectangle (3.61,0.9485);
\node[below,font=\scriptsize] at (3.35,-0.03) {L3};
\node[above,font=\scriptsize] at (3.35,0.9485) {0.948};
\node[font=\scriptsize] at (2.35,1.17) {Generated-tool Micro-F1};
\end{tikzpicture}
\end{minipage}
\hfill
\begin{minipage}{0.48\textwidth}
\centering
\begin{tikzpicture}[x=1.72cm,y=3.15cm]
\draw[->] (0,0) -- (4.8,0);
\draw[->] (0,0) -- (0,1.08);
\foreach \y in {0,0.25,0.5,0.75,1.0} {\draw (-0.04,\y) -- (0.04,\y) node[left,font=\scriptsize] {\y};}
\fill[teal!65!black] (0.69,0) rectangle (1.21,1.0000);
\node[below,font=\scriptsize] at (0.95,-0.03) {L1};
\node[above,font=\scriptsize] at (0.95,1.0000) {100.0\%};
\fill[teal!65!black] (1.89,0) rectangle (2.41,0.9000);
\node[below,font=\scriptsize] at (2.15,-0.03) {L2};
\node[above,font=\scriptsize] at (2.15,0.9000) {90.0\%};
\fill[teal!65!black] (3.09,0) rectangle (3.61,0.8000);
\node[below,font=\scriptsize] at (3.35,-0.03) {L3};
\node[above,font=\scriptsize] at (3.35,0.8000) {80.0\%};
\node[font=\scriptsize] at (2.35,1.17) {Live sandbox pass rate};
\end{tikzpicture}
\end{minipage}
\caption{End-to-end generation benchmark trends across L1 smoke, L2 realistic, and L3 adversarial suites. L2 has the largest gap between deterministic artifact quality and live sandbox success, which is consistent with more realistic file-processing edge cases.}
\label{fig:e2e-generation-graphs}
\end{figure*}

%% file: sections/10_governance.tex
Tool Forge assumes that generated code, model outputs, and imported tool
metadata are not intrinsically trusted. The primary threats include generated
code that performs unintended actions, generated tests that make live network
calls, hardcoded secrets, stale dependencies, incorrect API flows, imported MCP
tools with broad write capability, and agents that call tools outside their
intended session.

Several controls follow from this assumption. Credentials are referenced by
environment variable or credential mapping, not embedded in generated code.
Unit tests for external API tools are expected to mock network calls. Sandbox
validation requests real runtime values only when necessary. Router sessions
bind tool availability to a task and profile. Imported MCP tools begin in
pending review. Audit logs record metadata and argument names rather than raw
secret values.

This governance model matters because enterprise agent systems are judged not
only by whether they can act, but by whether they can explain and constrain
that action. When an agent calls a tool, the organization needs to know which
intent led to selection, which tool version was used, what validation evidence
existed, what credential alias was required, what profile permitted the call,
and what result was returned. Tool Forge makes these fields part of the
artifact and router state.

The model is not a replacement for OS-level sandboxing, network egress policy,
container isolation, RBAC, or human approval workflows. It is an application
control plane that can integrate with those lower-level controls. This
distinction is important for production deployments. A generated tool that can
move money, delete data, or post to customers should still require stronger
approval and monitoring than a local CSV transformation.

%% file: sections/11_limitations.tex
The current implementation has several limitations. First, documentation
grounding is constrained by the quality and crawlability of public API
documentation. Some providers distribute crucial flow information across
multiple pages, examples, and authentication guides. Generic endpoint extraction
can help, but it cannot fully replace provider-specific reasoning or human
review.

Second, the paper does not yet include a matched cross-system baseline against
other agent frameworks or code-generation products. This is deliberate rather
than an omission of a favorable result. A fair baseline would need identical
intents, runtime inputs, dependency policies, mock requirements, sandbox
execution, and artifact scoring. The current evaluation therefore emphasizes
reproducible within-system baselines: full-catalog exposure versus routed
exposure, and deterministic artifact checks versus live sandbox execution.

Third, local model mode improves privacy and experimentation but does not
guarantee high-quality synthesis. Small language models can struggle with long
contexts, JSON output validity, and complex API flows. Tool Forge mitigates
this through tiny generation and deterministic scaffolding, but the quality of
the generated core implementation still depends on model capability.

Fourth, Router retrieval remains an empirical system. The L3 benchmark shows
that negation and confusable tool families are still difficult. Better
retrieval features, explicit operation taxonomies, read/write classification,
policy constraints, and reranking are necessary for high-stakes settings.

Fifth, the generated-tool benchmark in this paper is a fixed 25-case probe,
not a full production certification. It covers L1 smoke, L2 realistic, and L3
adversarial local-tool tasks with a fixed model configuration, but it does not
cover hundreds of third-party APIs, long-running workflows, very large files,
or provider-specific authentication failures. The two sandbox failures in the
run are useful evidence that deterministic scoring and live execution should be
reported together rather than collapsed into a single quality number.

Sixth, release-time benchmarks can create a false sense of completeness if
their scope is not stated clearly. The present results are best understood as a
starting measurement for an open benchmark program. Stronger evidence would
include larger E2E suites, repeated runs across model providers and generation
recipes, live third-party API tasks with mocked and real validation modes,
human-authored baselines, and matched comparisons against other agent-tooling
systems under the same harness.

Finally, validation evidence is not proof. A sandbox can validate one path and
still miss production edge cases such as rate limits, permission boundaries,
large files, malformed provider responses, or provider API drift. Tool Forge
therefore should be viewed as a control-plane layer that makes evidence
explicit and enforceable, not as a theorem prover for arbitrary generated code.

%% file: sections/12_conclusion.tex
Agentic systems will not become reliable production software merely by adding
larger models or longer context windows. They require operational control
layers around the model. One of the most important layers is governed executable
capability: the ability to create, validate, catalog, route, and audit tools.

Tool Forge introduces a validation-carrying toolchain for this layer.
It treats a generated tool as a capsule containing intent, contract,
implementation, dependency policy, tests, documentation, validation evidence,
lifecycle state, and routing metadata. It treats the agent-facing tool context
as an intent-scoped projection over a governed catalog rather than a static
full-schema list.

The current implementation demonstrates promising results for token-efficient
tool routing and provides a practical open-source foundation for governed
intent-to-tool generation. The benchmark results also expose the next problems
clearly: adversarial routing, deeper API grounding, broader end-to-end
generation evaluation, stronger sandbox isolation, and additional language
targets. These are engineering and research directions, but they do not change
the core thesis. If agents are going to run real systems, tools must become
validation-carrying, governable, and context-efficient execution artifacts.